\documentclass[preprintnumbers, floatfix, twocolumn,
preprintnumbers, letterpaper, superscriptaddress,nofootinbib]{revtex4}
\usepackage{amsfonts}
\usepackage{amsmath}
\usepackage{amssymb,epsf}
\usepackage{latexsym}
\usepackage{graphicx,epsfig}
\usepackage{amssymb}
\usepackage{subfigure}
\usepackage[colorlinks=true,citecolor=blue,linkcolor=blue,urlcolor=black]{hyperref}

\begin{document}

\title{Traversable wormholes satisfying the weak energy condition\\
in third-order Lovelock gravity}
\author{Mahdi Kord Zangeneh}
\email{mkzangeneh@shirazu.ac.ir}
\affiliation{Physics Department and Biruni Observatory, Shiraz University, Shiraz 71454,
Iran}
\author{Francisco S. N. Lobo}
\email{fslobo@fc.ul.pt}
\affiliation{Instituto de Astrof\'isica e Ci\^encias do Espa\c{c}o, Universidade de
Lisboa, Faculdade de Ci\^encias, Campo Grande, PT1749-016 Lisboa, Portugal}
\author{Mohammad Hossein Dehghani}
\email{mhd@shirazu.ac.ir}
\affiliation{Physics Department and Biruni Observatory, Shiraz University, Shiraz 71454,
Iran}
\affiliation{Research Institute for Astronomy and Astrophysics of Maragha (RIAAM), P.O.
Box 55134-441, Maragha, Iran}
\date{\today }

\begin{abstract}
In this paper, we consider third order Lovelock gravity with a cosmological
constant term in an $n$-dimensional spacetime $\mathcal{M}^{4}\times 
\mathcal{K}^{n-4}$, where $\mathcal{K}^{n-4} $ is a constant curvature
space. We decompose the equations of motion to four and higher dimensional
ones and find wormhole solutions by considering a vacuum $\mathcal{K}^{n-4} $
space. Applying the latter constraint, we determine the second and third
order Lovelock coefficients and the cosmological constant in terms of
specific parameters of the model, such as the size of the extra dimensions.
Using the obtained Lovelock coefficients and $\Lambda$, we obtain the $4$%
-dimensional matter distribution threading the wormhole. Furthermore, by
considering the zero tidal force case and a specific equation of state,
given by $\rho =(\gamma p-\tau )/[\omega (1+\gamma )]$, we find the exact
solution for the shape function which represents both asymptotically flat
and non-flat wormhole solutions. We show explicitly that these wormhole
solutions in addition to traversibility satisfy the energy conditions for
suitable choices of parameters and that the existence of a limited
spherically symmetric traversable wormhole with normal matter in a $4 $%
-dimensional spacetime, implies a negative effective cosmological constant.
\end{abstract}

\maketitle

\section{Introduction}

Wormholes are tunnel-like objects, connecting two different regions of
spacetime, which include a throat-like region with a minimum radius. The
concept can be traced back as far as the 1916 paper by Flamm \cite{1}, where
he considered a \textquotedblleft tunnel-shaped\textquotedblright\ nature of
space near the Schwarzschild radius, which is possibly analogous to the
modern concept of the wormhole \textquotedblleft throat\textquotedblright .
However, one can consider the 1935 paper of Einstein and Rosen as the first
serious work on wormhole physics. The so-called Einstein-Rosen bridge was a
construction of an elementary particle model represented by a
\textquotedblleft bridge\textquotedblright\ connecting two identical sheets 
\cite{1}. Nevertheless, it was later shown that the Einstein-Rosen bridge is
unstable, as it collapses before a photon has had time to pass through \cite%
{3}. The term `wormhole' was coined for the first time in 1957 by John
Wheeler \cite{2} and a few wormhole solutions were investigated \cite{6288},
before the seminal Morris-Thorne paper in 1988, in which a static
traversable wormhole solution was presented for the first time. It was shown
that the energy-momentum distribution threading these solutions violated the
energy conditions, in particular, it violates the null energy condition
(NEC), and which has been denoted \textit{exotic matter} \cite{4,5}.
In fact, it was shown that the exoticity of the matter is a generic and
universal property of static wormholes within general relativity. This
analysis was presented taking into account the local geometry at the throat,
or in its vicinity, where the assumption of asymptotic flatness is not a
necessary requirement \cite{13,14,NEC0}.

Recently, the late-time accelerated expansion of the Universe \cite{dark}
has also motivated research in wormhole physics, due to the possible
presence of an exotic cosmic fluid responsible for this cosmic acceleration,
i.e., phantom energy, that violates the null energy condition \cite%
{phantomWH}. Observational features of wormholes physics have also been
extensively analysed in the literature \cite{Harko:2008vy}. Relative to the
energy condition violations, some attempts have been done in the literature
to construct wormholes with normal matter \cite{normal}. Indeed, as the
violation of the energy conditions is a particularly problematic issue \cite%
{Lobo:2004wq}, it is necessary to minimize its usage. In fact, to this aim,
wormholes have been substantially examined from different points of view in
the literature \cite%
{diffpo,diffpo0,diffpo4,diffpo2,diffpo1,diffpo3,18,modgravWH}. In \cite%
{brane1}, a class of static and spherically symmetric solutions in a vacuum
brane was obtained where it was shown that bulk Weyl effects support the
wormhole. In fact, as it is the effective total stress energy tensor that
violates the energy conditions, one may impose that the energy-momentum
tensor confined on the brane, threading the wormhole, satisfies the NEC.
Furthermore, the analysis carried out in \cite{brane1} was generalized in 
\cite{brane2}, by showing that in addition to the nonlocal corrections from
the Weyl curvature in the bulk, the local high-energy bulk effects may
induce a NEC violating signature on the brane. Thus, braneworld gravity
seems to provide a natural scenario for the existence of traversable
wormholes.

From the beginning of introducing higher dimensional spacetimes, physicists
faced questions on the size of these non-observable extra dimensions. The
problem was first solved by compactifying the extra dimension on small
enough regions, as the non-observability of the extra dimensions was due to
their smallness. This idea has recently been substituted by another in that
higher dimensions can be large but they are not observable as no observable
matter and fields can exist there, and it is only gravity that propagates in
these extra dimensions \cite{Maartens:2003tw}. The latter approach is
designated by the \textquotedblleft brane-world\textquotedblright\ scenario 
\cite{brane0}. The general formalism of the brane-world is to consider the
metric of the observable world on a \textquotedblleft
brane\textquotedblright , which is embedded in a higher dimensional space
denoted by the \textquotedblleft bulk\textquotedblright . In this paper,
although we do not use the typical formalism of the brane-world scenario, we
apply its idea to obtain wormhole solution. We consider a theory consisting
of third order Lovelock gravity \cite{lovee} in addition to a cosmological
constant. The constants of the theory are fixed so that there is no
energy-momentum tensor except for the observable world. Applying this, we
first present both asymptotically flat and non-flat novel wormhole solutions
in third-order Lovelock gravity, and investigate the physical properties of
these solutions, showing that it is possible that these traversable
wormholes may be constructed from normal matter.

In fact, it is interesting to note that it has recently been shown that
exact traversable wormholes in $(3+1)$-dimensional general relativity can be
constructed, in which the only \textquotedblleft exotic
source\textquotedblright\ is a negative cosmological constant (which can
hardly be defined as exotic) \cite{Ayon-Beato:2015eca}. Based on the
techniques developed in Refs. \cite{technique}, the authors found the first
self-gravitating, analytic and globally regular Skyrmion solution of the
Einstein-Skyrme system, in the presence of a cosmological constant. More
specifically, the equations admit analytic bouncing cosmological solutions
in which the universe contracts to a minimum non-vanishing size, and then
expands. A non-trivial byproduct of this solution is that a minor
modification of the construction gives rise to a family of stationary,
regular configurations in general relativity with a negative cosmological
constant supported by an SU(2) nonlinear sigma model. These solutions
represent traversable AdS wormholes with a NUT parameter in which the only
\textquotedblleft exotic matter\textquotedblright\ required for their
construction is a negative cosmological constant. Thus, for instance, one
can have that a cloud of interacting Pions (described with the usual
nonlinear sigma model) in $(3+1)$-dimensional general relativity can support
a traversable AdS wormhole with NUT charge. Moreover, this wormhole is
likely to be stable as the Pions are in a configuration with a non-trivial
topological charge and so they cannot collapse to the trivial Pions vacuum.
As a last remark, at a first glance, if one deals with the NUT parameter as
it is usually done, then there are closed time-like curves. However, it was
recently found that closed time-like curves can be avoided in spacetimes
with a NUT parameter when these are regular \cite{Clement:2015cxa}, as is
the case with the solutions found in \cite{Ayon-Beato:2015eca}, so that one
may, in principle, get rid of the closed time-like curves in the latter
case, as well.

This paper is outlined in the following manner: In Section \ref{FE}, we
present the field equations in third order Lovelock gravity in the presence
of a cosmological constant term, which will be used in the subsequent
sections. In Section \ref{Bra}, by considering a $4$-dimensional spacetime $%
\mathcal{M}^{4}$ which is a part of an $n$-dimensional $\mathcal{M}%
^{4}\times \mathcal{K}^{n-4}$ spacetime where $\mathcal{K}^{n-4}$ is a
constant curvature space which we consider empty, we decompose the equations
of motion to four and higher dimensional ones, and consequently write out
the general four-dimensional gravitatinal field equations. In Section \ref%
{sec:phys}, by imposing a constant redshift function and considering a
specific equation of state, we find specific wormhole solutions, i.e.,
asymptotically flat and general wormhole spacetimes, and show that these
solutions satisfy the energy conditions and are traversable. Furthermore,
for the non-asymptotically flat spacetimes, these interior geometries are
matched to an exterior spherically-symmetric spacetime, in an Ads Universe.

\section{Third order Lovelock gravity: Field Equations}

\label{FE}

The generalized form of the gravitational field equation which maintains the
properties of the Einstein equation is the Lovelock equation. Thus, we
consider third order Lovelock gravity \cite{lovee} with a cosmological
constant term, where the gravitational field equations are given by 
\begin{equation}
\mathcal{G}_{\mu \nu }=G_{\mu \nu }^{(1)}+\alpha _{2}G_{\mu \nu
}^{(2)}+\alpha _{3}G_{\mu \nu }^{(3)}+\Lambda g_{\mu \nu }=\kappa \mathcal{T}%
_{\mu \nu },  \label{Geq}
\end{equation}%
where $G_{\mu \nu }^{(1)}$ is the Einstein tensor, the constants $\alpha
_{i} $'s are the Lovelock coefficients, $g_{\mu \nu }$ is the metric, the
coupling constant of matter and gravity $\kappa $ is set equal to $1$ ($%
\kappa =1$), and $\mathcal{T}_{\mu \nu }$ is the energy-momentum tensor. $%
G_{\mu \nu }^{(2)}$ and $G_{\mu \nu }^{(3)}$ are the second and third order
Lovelock tensors provided by 
\begin{eqnarray}
G_{\mu \nu }^{(2)} &=&2\big(R_{\mu \sigma \kappa \tau }R_{\nu }^{%
\phantom{\nu}\sigma \kappa \tau }-2R_{\mu \rho \nu \sigma }R^{\rho \sigma
}-2R_{\mu \sigma }R_{\phantom{\sigma}\nu }^{\sigma }  \notag \\
&&+RR_{\mu \nu }\big)-\frac{1}{2}\mathcal{L}_{2}g_{\mu \nu },  \label{Love2}
\end{eqnarray}%
and 
\begin{eqnarray}
G_{\mu \nu }^{(3)} &=&-3(4R^{\tau \rho \sigma \kappa }R_{\sigma \kappa
\lambda \rho }R_{\phantom{\lambda }{\nu \tau \mu}}^{\lambda }-8R_{%
\phantom{\tau \rho}{\lambda \sigma}}^{\tau \rho }R_{\phantom{\sigma
\kappa}{\tau \mu}}^{\sigma \kappa }R_{\phantom{\lambda }{\nu \rho \kappa}%
}^{\lambda }  \notag \\
&&+2R_{\nu }^{\phantom{\nu}{\tau \sigma \kappa}}R_{\sigma \kappa \lambda
\rho }R_{\phantom{\lambda \rho}{\tau \mu}}^{\lambda \rho }-R^{\tau \rho
\sigma \kappa }R_{\sigma \kappa \tau \rho }R_{\nu \mu }  \notag \\
&&+8R_{\phantom{\tau}{\nu \sigma \rho}}^{\tau }R_{\phantom{\sigma
\kappa}{\tau \mu}}^{\sigma \kappa }R_{\phantom{\rho}\kappa }^{\rho }+8R_{%
\phantom {\sigma}{\nu \tau \kappa}}^{\sigma }R_{\phantom {\tau \rho}{\sigma
\mu}}^{\tau \rho }R_{\phantom{\kappa}{\rho}}^{\kappa }  \notag \\
&&+4R_{\nu }^{\phantom{\nu}{\tau \sigma \kappa}}R_{\sigma \kappa \mu \rho
}R_{\phantom{\rho}{\tau}}^{\rho }-4R_{\nu }^{\phantom{\nu}{\tau \sigma
\kappa }}R_{\sigma \kappa \tau \rho }R_{\phantom{\rho}{\mu}}^{\rho }  \notag
\\
&&+4R^{\tau \rho \sigma \kappa }R_{\sigma \kappa \tau \mu }R_{\nu \rho
}+2RR_{\nu }^{\phantom{\nu}{\kappa \tau \rho}}R_{\tau \rho \kappa \mu } 
\notag \\
&&+8R_{\phantom{\tau}{\nu \mu \rho }}^{\tau }R_{\phantom{\rho}{\sigma}%
}^{\rho }R_{\phantom{\sigma}{\tau}}^{\sigma }-8R_{\phantom{\sigma}{\nu \tau
\rho }}^{\sigma }R_{\phantom{\tau}{\sigma}}^{\tau }R_{\mu }^{\rho }  \notag
\\
&&-8R_{\phantom{\tau }{\sigma \mu}}^{\tau \rho }R_{\phantom{\sigma}{\tau }%
}^{\sigma }R_{\nu \rho }-4RR_{\phantom{\tau}{\nu \mu \rho }}^{\tau }R_{%
\phantom{\rho}\tau }^{\rho }  \notag \\
&&+4R^{\tau \rho }R_{\rho \tau }R_{\nu \mu }-8R_{\phantom{\tau}{\nu}}^{\tau
}R_{\tau \rho }R_{\phantom{\rho}{\mu}}^{\rho }  \notag \\
&&+4RR_{\nu \rho }R_{\phantom{\rho}{\mu }}^{\rho }-R^{2}R_{\nu \mu })-\frac{1%
}{2}\mathcal{L}_{3}g_{\mu \nu },  \label{Love3}
\end{eqnarray}%
respectively, where 
\begin{equation}
\mathcal{L}_{2}=R_{\mu \nu \gamma \delta }R^{\mu \nu \gamma \delta }-4R_{\mu
\nu }R^{\mu \nu }+R^{2}\,,
\end{equation}%
is the Gauss-Bonnet Lagrangian and%
\begin{eqnarray}
\mathcal{L}_{3} &=&2R^{\mu \nu \sigma \kappa }R_{\sigma \kappa \rho \tau }R_{%
\phantom{\rho \tau }{\mu \nu }}^{\rho \tau }+8R_{\phantom{\mu \nu}{\sigma
\rho}}^{\mu \nu }R_{\phantom {\sigma \kappa} {\nu \tau}}^{\sigma \kappa }R_{%
\phantom{\rho \tau}{ \mu \kappa}}^{\rho \tau }  \notag \\
&&+24R^{\mu \nu \sigma \kappa }R_{\sigma \kappa \nu \rho }R_{%
\phantom{\rho}{\mu}}^{\rho }+3RR^{\mu \nu \sigma \kappa }R_{\sigma \kappa
\mu \nu }  \notag \\
&&+24R^{\mu \nu \sigma \kappa }R_{\sigma \mu }R_{\kappa \nu }+16R^{\mu \nu
}R_{\nu \sigma }R_{\phantom{\sigma}{\mu}}^{\sigma }  \notag \\
&&-12RR^{\mu \nu }R_{\mu \nu }+R^{3}\,,
\end{eqnarray}%
is the third order Lovelock Lagrangian.

It is worthwhile to mention that due to the absence of derivatives of the
curvatures in Eq. (\ref{Geq}), it does not include the derivatives of the
metric higher than two. It is also remarkable to note that the dimension of
the spacetime should be equal or larger than seven so that all the terms in (%
\ref{Geq}) contribute in the field equation\cite{lovee}.

\section{Wormhole Solutions: Metric and field equations}

\label{Bra}


In this section, we aim to present the field equations of a $4$-dimensional
Morris and Thorne wormhole which is a part of an $n$-dimensioanl spacetime
in third order Lovelock gravity. More specifically, we will investigate the
properties and characteristics of wormholes in a $4$-dimensional spacetime $%
\mathcal{M}^{4}$ with the following metric 
\begin{equation}
g_{ab}dx^{a}dx^{b}=-e^{2\phi (r)}dt^{2}+\left( 1-\frac{b(r)}{r}\right)
^{-1}dr^{2}+r^{2}d\Sigma _{2(\hat{k})}^{2},  \label{metric}
\end{equation}%
where $a,b=0,1,2,3$, and $d\Sigma _{2(\hat{k})}^{2}$ represents the line
element of a $2$-dimensional hypersurface with constant curvature $2\hat{k}$%
, with $\hat{k}=0,\pm 1$. The functions $\phi (r)$ and $b(r)$ are denoted
the redshift function and shape function, respectively. We consider metric (%
\ref{metric}) as part of an $n$-dimensional spacetime $\mathcal{M}^{4}\times 
\mathcal{K}^{n-4}$ with the metric%
\begin{equation}
ds^{2}=g_{ab}dx^{a}dx^{b}+r_{0}^{2}\gamma _{ij}d\theta ^{i}d\theta ^{j}.
\label{met}
\end{equation}%
where $i,j=4...(n-1)$, $r_{0}$ is a constant and $\gamma _{ij}$ is the
metric of $(n-4)$-dimensional $\mathcal{K}^{n-4}$ space which has constant
curvature. The constant curvature of $\mathcal{K}^{n-4}$ is $(n-4)(n-5)k$,
where $k=0,\pm 1$.

The metric (\ref{metric}) represents a traversable wormhole provided the
function $\phi (r)$ is finite everywhere and the shape function $b(r)$
satisfies the following two conditions:%
\begin{eqnarray}
d(r) &\equiv &b(r)-r\leq 0\text{ \ \ \ \ \ \ \ }r\in \lbrack a_{0},a),
\label{Cond1} \\
F(r) &\equiv &rb^{\prime }-b(r)<0\text{ \ \ \ \ \ }r\in (a_{0},a),
\label{Cond2}
\end{eqnarray}%
respectively, where $a_{0}$ is the throat radius and the prime denotes a
derivative with respect to the radial coordinate $r$. Note that the equality
in Eq. (\ref{Cond1}) only occurs at $r=a_{0}$, i.e., for $b(a_{0})=a_{0}$.
The parameter $a$ could, in principle, be as large as $+\infty $ but it is
typically set to be finite and resides at the throat neighbourhood. The
first condition is due to the fact that the proper radial distance $l(r)=\pm
\int_{a_{0}}^{r} \left[1-b(r)/r \right]^{-1/2} \,dr$, should be real and
finite for $r>a_{0}$. The second condition arises from the flaring-out
condition \cite{4}.

The mathematical analysis and the physical interpretation will be simplified
using a set of orthonormal basis vectors%
\begin{gather}
\mathbf{e}_{\hat{{t}}}=e^{-\phi (r)}\frac{\partial }{\partial t},\text{ \ \
\ }\mathbf{e}_{\hat{r}}=\left( 1-\frac{b(r)}{r}\right) ^{1/2}\frac{\partial 
}{\partial r},  \notag \\
\mathbf{e}_{\hat{\theta}}=r^{-1}\frac{\partial }{\partial \theta },\text{ \
\ \ }\mathbf{e}_{\hat{\varphi}}=\left\{ 
\begin{array}{l}
\left( r\sin \theta \right) ^{-1}\frac{\partial }{\partial \varphi }\text{ \
\ \ for \ \ }\hat{k}=1 \\ 
\frac{1}{r}\frac{\partial }{\partial \varphi }\text{ \ \ \ \ \ \ \ \ \ \ \ \
\ \ \ for \ \ }\hat{k}=0 \\ 
\left( r\sinh \theta \right) ^{-1}\frac{\partial }{\partial \varphi }\text{
\ for \ \ }\hat{k}=-1%
\end{array}%
\right. ,  \notag \\
\mathbf{e}_{\hat{\imath}\hat{\jmath}}=(r_{0}\sqrt{\gamma _{ij}})^{-1}\frac{%
\partial }{\partial \theta ^{i}}.  \label{bas}
\end{gather}

Thus, Eq. (\ref{Geq}) finally provides the following decomposition 
\begin{widetext}
\begin{gather}
\mathcal{G}_{\hat{a}\hat{b}}=\left\{ -\frac{(n-4)(n-5)k}{2r_{0}^{2}}\left[ 1+%
\frac{k\alpha _{2}(n-6)(n-7)}{r_{0}^{2}}+\frac{\alpha
_{3}(n-6)(n-7)(n-8)(n-9)}{r_{0}^{4}}\right] +\Lambda \right\} g_{\hat{a}\hat{%
b}}  \notag \\
+\left\{ 1+\frac{2k\alpha _{2}(n-4)(n-5)}{r_{0}^{2}}+%
\frac{3\alpha _{3}k^{2}(n-4)(n-5)(n-6)(n-7)}{r_{0}^{4}}\right\} \tilde{G}_{%
\hat{a}\hat{b}},  \label{Gab}
\end{gather}%
\begin{gather}
\mathcal{G}_{\hat{\imath}\hat{\jmath}\ }=-\frac{1}{2}\left\{ \frac{%
(n-5)(n-6)k}{r_{0}^{2}}\left[ 1+\frac{k\alpha _{2}(n-7)(n-8)}{r_{0}^{2}}+%
\frac{\alpha _{3}(n-7)(n-8)(n-9)(n-10)}{r_{0}^{4}}-\frac{2\Lambda r_{0}^{2}}{%
(n-5)(n-6)k}\right] \right.   \notag \\
\hspace{1cm}\left. +\left[ 1+\frac{2k\alpha _{2}(n-5)(n-6)}{r_{0}^{2}}+\frac{%
3k^{2}\alpha _{3}(n-5)(n-6)(n-7)(n-8)}{r_{0}^{4}}\right] \right. \tilde{R} 
\notag \\
\hspace{1cm}\left. +\left[ \alpha _{2}+\frac{3k\alpha _{3}(n-5)(n-6)}{%
r_{0}^{2}}\right] \mathcal{\tilde{L}}_{2}\right\} g_{\hat{\imath}\hat{\jmath}%
},  \label{Gij}
\end{gather}%
\end{widetext}
where the superscripts ``tilde'' represent quantities on $\mathcal{M}^{4}$.

In the following, we construct a $4$-dimensional wormhole solution where $%
\mathcal{K}^{n-4}$ space is empty. In general, $\tilde{R}$ and $\mathcal{%
\tilde{L}}_{2}$ do not vanish, however, if the three brackets in Eq. (\ref%
{Gij}) vanish, we have $\mathcal{G}_{\hat{\imath}\hat{\jmath}}=0$ . Thus,
setting the quantities in the three square brackets in Eq. (\ref{Gij}) equal
to zero, for $k\neq 0$, one obtains for the specific cases of $k=\pm 1$, the
follwoing relations 
\begin{equation}
\alpha _{2}=\frac{-{r_{0}}^{2}}{\Xi k},  \label{A2}
\end{equation}%
\begin{equation}
\alpha _{3}=\frac{{r_{0}}^{4}}{3(n-5)(n-6)\Xi },  \label{A3}
\end{equation}%
\begin{eqnarray}
\Lambda  &=&\frac{k}{6\Xi {r_{0}}^{2}}\left\{
6(n-5)^{2}(n-6)^{2}-(n-7)(n-8)\right.   \notag \\
&&\left. \times \left[ 6(n-5)(n-6)-(n-9)(n-10)\right] \right\} ,  \label{cc0}
\end{eqnarray}%
where, for notational simplicity, we have defined 
\begin{equation}
\Xi =2(n-5)(n-6)-(n-7)(n-8)\,.
\end{equation}%
Note that in order to have a finite value for $\alpha _{3}$, we need $n\geq 7
$ and therefore $\Xi >0$. Now inserting the above values of $\Lambda $, $%
\alpha _{2}$ and $\alpha _{3}$ in Eq. (\ref{Gab}), one obtains%
\begin{equation}
\mathcal{G}_{\hat{a}\hat{b}\ }=-\frac{8}{\Xi }\left( \tilde{G}_{\hat{a}\hat{b%
}}+g_{\hat{a}\hat{b}}\Lambda _{\mathrm{eff}}\right) ,  \label{Gab2}
\end{equation}%
where\qquad \qquad\ 
\begin{equation}
\Lambda _{\mathrm{eff}}=\frac{(2n-13)k}{r_{0}^{2}}.  \label{efcc}
\end{equation}%
One should note that $\mathcal{G}_{\hat{a}\hat{b}\ }\neq 0$ even if $\tilde{G%
}_{\hat{a}\hat{b}}=0$. Therefore, one may consider the effects of higher
curvature terms of Lovelock gravity in $4$-dimensional solutions as a
cosmological constant term with an effective constant $\Lambda _{\mathrm{eff}%
}$, given by Eq. (\ref{efcc}). Furthermore, $\Lambda _{\mathrm{eff}}$ can be
positive or negative for $k=1$ or $k=-1$, respectively (note that $n\geq 7$).

Next, we calculate the energy density, radial tension and pressure
associated to the matter of the $4$-dimensional manifold $\mathcal{M}^{4}$,
in order to study the properties and characterisitics of the solution. Using
the orthonormal basis (\ref{bas}), the components of the energy-momentum
tensor $\mathcal{T}_{\hat{a}\hat{b}}$, carry a simple physical
interpretation, i.e.,%
\begin{gather}
\mathcal{T}_{_{\hat{t}\hat{t}}}=\mathcal{G}_{_{\hat{t}\hat{t}}\ }\equiv \rho
(r),\text{ \ \ }\mathcal{T}_{_{\hat{r}\hat{r}}}=\mathcal{G}_{_{\hat{r}\hat{r}%
}}\equiv -\tau (r),  \notag \\
\mathcal{T}_{\hat{\theta}\hat{\theta}}=\mathcal{T}_{\hat{\varphi}\hat{\varphi%
}}=\mathcal{G}_{\hat{\theta}\hat{\theta}}\equiv p(r),
\end{gather}%
where $\rho $ is the energy density, $\tau $ is the radial tension, and $p$
is the pressure measured in the tangential directions orthogonal to the
radial direction. The radial tension is $\tau =-p_{r}$, where $p_{r}$ is the
radial pressure. Calculating the components of $\mathcal{G}_{\hat{a}\hat{b}\
}$ for the metric (\ref{metric}), we obtain the following energy-momentum
profile 
\begin{equation}
\rho (r)=-\frac{8}{\Xi }\left\{ \frac{b^{\prime }}{r^{2}}+\frac{\hat{k}-1}{%
r^{2}}-\Lambda _{\mathrm{eff}}\right\} ,  \label{rhor}
\end{equation}%
\begin{eqnarray}
\tau (r) &=&-\frac{8}{\Xi }\left\{ \frac{b}{r^{3}}-2\left( 1-\frac{b}{r}%
\right) \frac{\phi ^{\prime }}{r}+\frac{\hat{k}-1}{r^{2}}\,-\Lambda _{%
\mathrm{eff}}\right\} ,  \notag \\
&&  \label{rhot}
\end{eqnarray}%
\begin{eqnarray}
p(r) &=&-\frac{8}{\Xi }\left\{ \left( 1-\frac{b}{r}\right) \left[ \phi
^{\prime \prime }+(\phi ^{\prime 2}-\frac{b^{\prime }r-b}{2r(r-b)}\phi
^{\prime }\right. \right.  \notag \\
&&\left. \left. -\frac{b^{\prime }r-b}{2r^{2}(r-b)}+\frac{\phi ^{\prime }}{r}%
\right] +\Lambda _{\mathrm{eff}}\,\right\} .  \label{rhott}
\end{eqnarray}%
Applying the conservation of the energy momentum tensor $\mathcal{T}_{\text{
\ \ };\hat{\nu}}^{\hat{\mu}\hat{\nu}}$ for $\hat{\mu}=r$, the relativistic
Euler equation is given by 
\begin{equation}
\tau ^{\prime }+(\tau -\rho )\phi ^{\prime }+\frac{2}{r}(p+\tau )=0.
\label{Euler Eq}
\end{equation}

In the next section, we study the physical properties of wormhole solutions
by applying an equation of state between components of energy-momentum
tensor.

\section{Physical Properties of the wormhole solutions}

\label{sec:phys}

In this section we study the physical properties of the wormhole solutions,
by considering a specific class of solutions corresponding to the choice of $%
\phi (r)=\text{const}=\Phi $. In order to solve Eqs. (\ref{rhor})-(\ref%
{rhott}) for $b(r)$, we assume the equation of state (EOS) given by $\rho
=(\gamma p-\tau )/[\omega (1+\gamma )]$\footnote{%
It is worth noting that for an isotropic matter i.e. $-\tau =p$, the EOS
reduces to the well-known isotropic EOS $p=\omega \rho$.} \cite{EOS}, so
that one finally obtains the following solution 
\begin{equation}
b(r)=\left( \hat{k}-a_{0}^{2}\beta \right) a_{0}^{1-\eta }r^{\eta }+\beta
r^{3}-(\hat{k}-1)r,  \label{shape}
\end{equation}%
where%
\begin{equation*}
\beta =\frac{(\omega +1)(2n-13)k}{r_{0}^{2}(3\omega +1)}, \qquad \eta =\frac{%
\gamma -2}{2(1+\gamma )\omega +\gamma }.
\end{equation*}%
It is easy to check that Eq. (\ref{shape}) satisfies the condition $%
b(a_{0})=a_{0}$. One can see from (\ref{shape}) that the wormhole solution
is asymptotically flat, i.e., $b(r)/r\rightarrow 0$ as $r\rightarrow \infty $%
, provided that $\beta =0$, $\hat{k}=1$ and $\eta <1$. Since $k\neq 0$ [we
refer the reader to the discussion above Eq. (\ref{A2})], $\beta =0$ is
equivalent to $\omega =-1$ and therefore $\eta $ reduces to $(2-\gamma
)/(2+\gamma )$ for this case. Thus, the condition $\eta <1$ for the
asymptotically flat solution is satisfied provided $\gamma >0$ or $\gamma
<-2 $. With the shape function (\ref{shape}) in hand, we now investigate the
energy conditions and the physical properties of the wormhole, for both the
asymptotically flat and general cases.

\subsection{Energy conditions}

\label{WEC}

Here, we analyse the energy conditions for the energy-momentum tensor
profile described by the system of equations (\ref{rhor})-(\ref{rhott}). In
particular, we consider the weak energy condition (WEC), which states that $%
\mathcal{T}_{\mu \nu}u^{\mu }u^{\nu }\geq 0$ where $u^{\mu }$ is the
timelike velocity of the observer. In terms of the non-zero components of
the diagonal energy-momentum tensor, we have the following inequalities: $%
\rho \geq 0$, $\rho -\tau \geq 0$ and $\rho +p\geq 0$. Note that the last
two inequalities are referred to as the null energy condition (NEC).

\subsubsection{Asymptotically flat case}

In the asymptotically flat case where $\beta =0$ (or $\omega =-1$), $\hat{k}%
=1$ and $\eta <1$, Eqs. (\ref{rhor})-(\ref{rhott}) and solution (\ref{shape}%
) yield the following relations 
\begin{equation}
\rho =\frac{8}{\Xi }\left( -\eta a_{0}^{1-\eta }r^{\eta -3}+\frac{(2n-13)k}{%
r_{0}^{2}}\right) \geq 0,  \label{rhoflat}
\end{equation}
\begin{equation}
\rho -\tau =\frac{8}{\Xi }(1-\eta )a_{0}^{1-\eta }r^{\eta -3}\geq 0,
\label{rhotflat}
\end{equation}
\begin{equation}
\rho +p=-\frac{4}{\Xi }(1+\eta )a_{0}^{1-\eta }r^{\eta -3}\geq 0.
\label{rhopflat}
\end{equation}
For $\eta \leq -1$, the inequalities (\ref{rhotflat}) and (\ref{rhopflat})
are readily satisfied, and consequently the NEC is also obeyed (note that $%
\Xi >0$). One can also see that the inequality (\ref{rhoflat}) is satisfied
for suitable choices of the parameters. For instance, for $\eta \leq 0$ and $%
k=1$, we have $\rho \geq 0$ (note that $k\neq 0$ and $n\geq 7$; see
discussions above Eq. (\ref{A2}) and below Eq. (\ref{cc0})). In concluding,
for the asymptotically flat case, the WEC is satisfied for suitable choice
of parameters, for example, $\eta \leq -1$ and $k=1$.

\subsubsection{General case}

For the general case, by taking into account Eqs. (\ref{rhor})-(\ref{rhott})
and (\ref{shape}), one arrives at the following inequalities 
\begin{equation}
\rho =\frac{8}{\Xi }\left[ (a_{0}^{2}\beta -\hat{k})\eta a_{0}^{1-\eta
}r^{\eta -3}-\frac{2\beta }{1+\omega }\right] \geq 0,  \label{rho}
\end{equation}%
\begin{equation}
\rho -\tau =\frac{8}{\Xi }\left[ (\hat{k}-a_{0}^{2}\beta )(1-\eta
)a_{0}^{1-\eta }r^{\eta -3}-2\beta \right] \geq 0,  \label{rhotau}
\end{equation}%
\begin{equation}
\rho +p=\frac{8}{\Xi }\left[ \frac{1}{2}(a_{0}^{2}\beta -\hat{k})(1+\eta
)a_{0}^{1-\eta }r^{\eta -3}-2\beta \right] \geq 0,  \label{rhop}
\end{equation}%
from which it is clear that satisfying the WEC relies on the values of $r$, $%
a_{0}$ and $r_{0}$, in general. However, independently of the specific
choice of the latter parameters, i.e., $r$, $a_{0}$ and $r_{0}$, one finds
that the WEC is satisfied for $\hat{k}=0,1$ provided that 
\begin{equation}
\beta <0,\qquad \text{and}\qquad \eta \leq -1.  \label{wec}
\end{equation}%
These inequalities can be expressed in terms of $\omega $\ and $\gamma $\ as
follows:

\begin{enumerate}
\item $k=-1$: In this case, the WEC is satisfied provided%
\begin{eqnarray}
\text{a. }\omega &<&-1\text{ \ \ and \ }\frac{1-\omega }{1+\omega }\leq
\gamma <2  \notag \\
\text{b. }\omega &>&-\frac{1}{3}\text{ and \ }\gamma \leq \frac{1-\omega }{%
1+\omega }.  \label{wec1}
\end{eqnarray}

\item $k=1$: In this case, inequality (\ref{wec}) is satisfied provided%
\begin{equation}
-1<\omega <-\frac{1}{3}\text{ \ and \ }\gamma \geq \frac{1-\omega }{1+\omega 
}.  \label{wec2}
\end{equation}
\end{enumerate}

\subsection{Conditions imposed on the shape function}

\label{trav}

In this section, we analyse the conditions (\ref{Cond1}) and (\ref{Cond2})
for both the asymptotically flat and general wormhole solutions considered
above.

\subsubsection{Asymptotically flat case}

The conditions (\ref{Cond1}) and (\ref{Cond2}) for the asymptotically flat
solution [$\beta =0$ (or $\omega =-1$), $\hat{k}=1$ and $\eta <1$] reduce to
the following 
\begin{eqnarray}
d(r) &=&a_{0}^{1-\eta }r^{\eta }-r\leq 0,  \label{d1} \\
F(r) &=&(\eta -1)a_{0}^{1-\eta }r^{\eta }<0.  \label{F1}
\end{eqnarray}%
Note that the inequality (\ref{F1}) is readily satisfied. In order to
analyse (\ref{d1}), one can rewrite it as $\left( a_{0}/r\right) ^{1-\eta
}\leq 1$ which is obviously satisfied since $r\geq a_{0}$ (note that $1-\eta
>0$ for the asymptotically flat solution).

\subsubsection{General case}

\begin{figure*}[tbp]
\centering{\includegraphics[width=.4\textwidth]{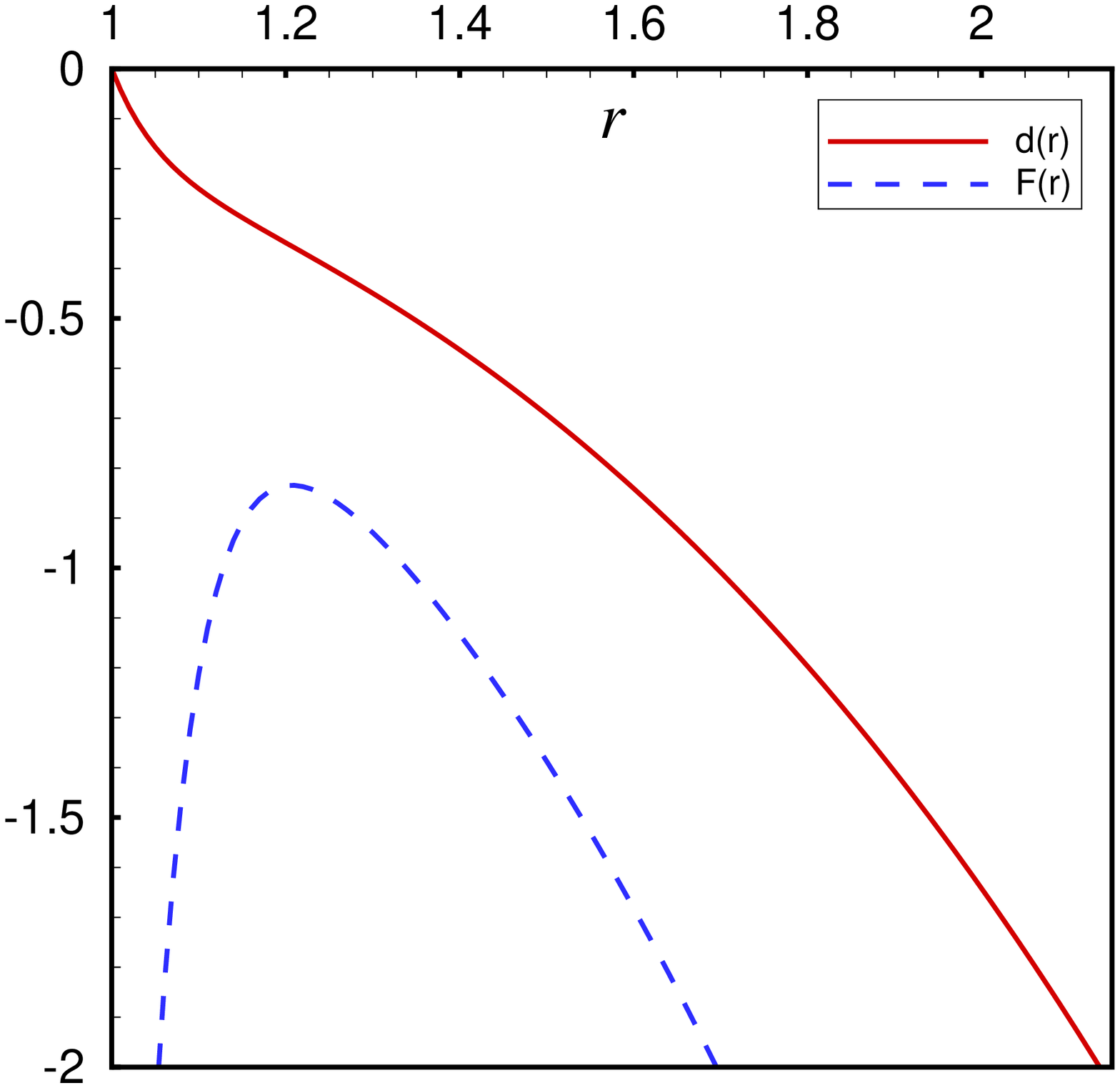}\qquad } {%
~~~~~~~~~~~~~} {\includegraphics[width=.4\textwidth]{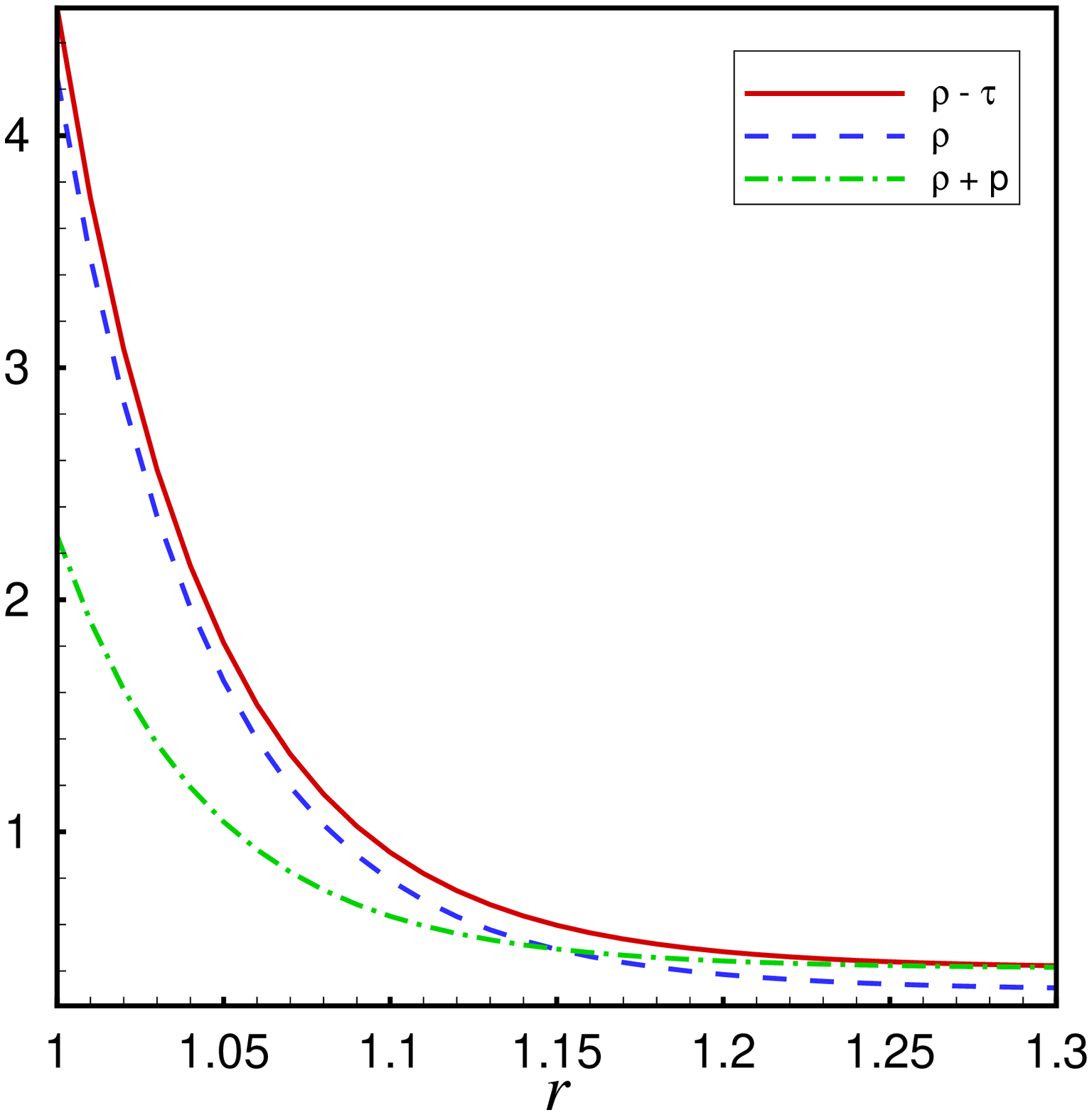}}
\caption{Left: $d(r)$ and $F(r)$ versus $r$; Right: $\protect\rho $, $%
\protect\rho -\protect\tau $ and $\protect\rho +p$ versus $r$ for $n=8$, $%
a_{0}=1$, $r_{0}=0.001$, $\protect\omega =0.3$, $k=-1$, $\protect\gamma %
=-0.3 $ and $\hat{k}=1$. The scale of the vertical axes of both figures is
divided by $10^{7}$. Note that the local maximum of the $F(r)$ corresponds
to the change of concavity of the $d(r)$ or similarly $b(r)$.}
\label{f1}
\end{figure*}
\begin{figure*}[tbp]
\centering{\includegraphics[width=.4\textwidth]{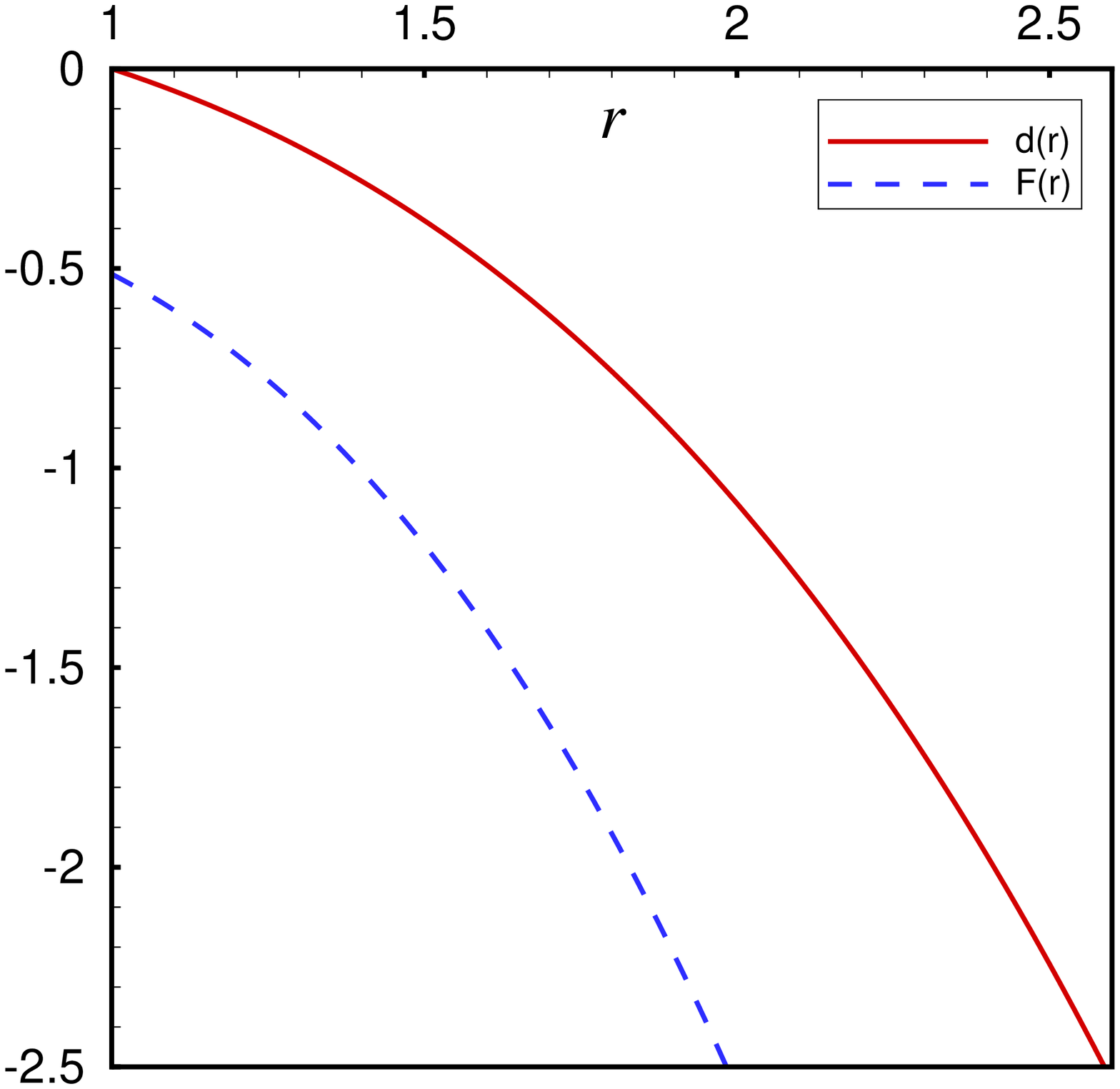}\qquad } {%
~~~~~~~~~~~~~} {\includegraphics[width=.4\textwidth]{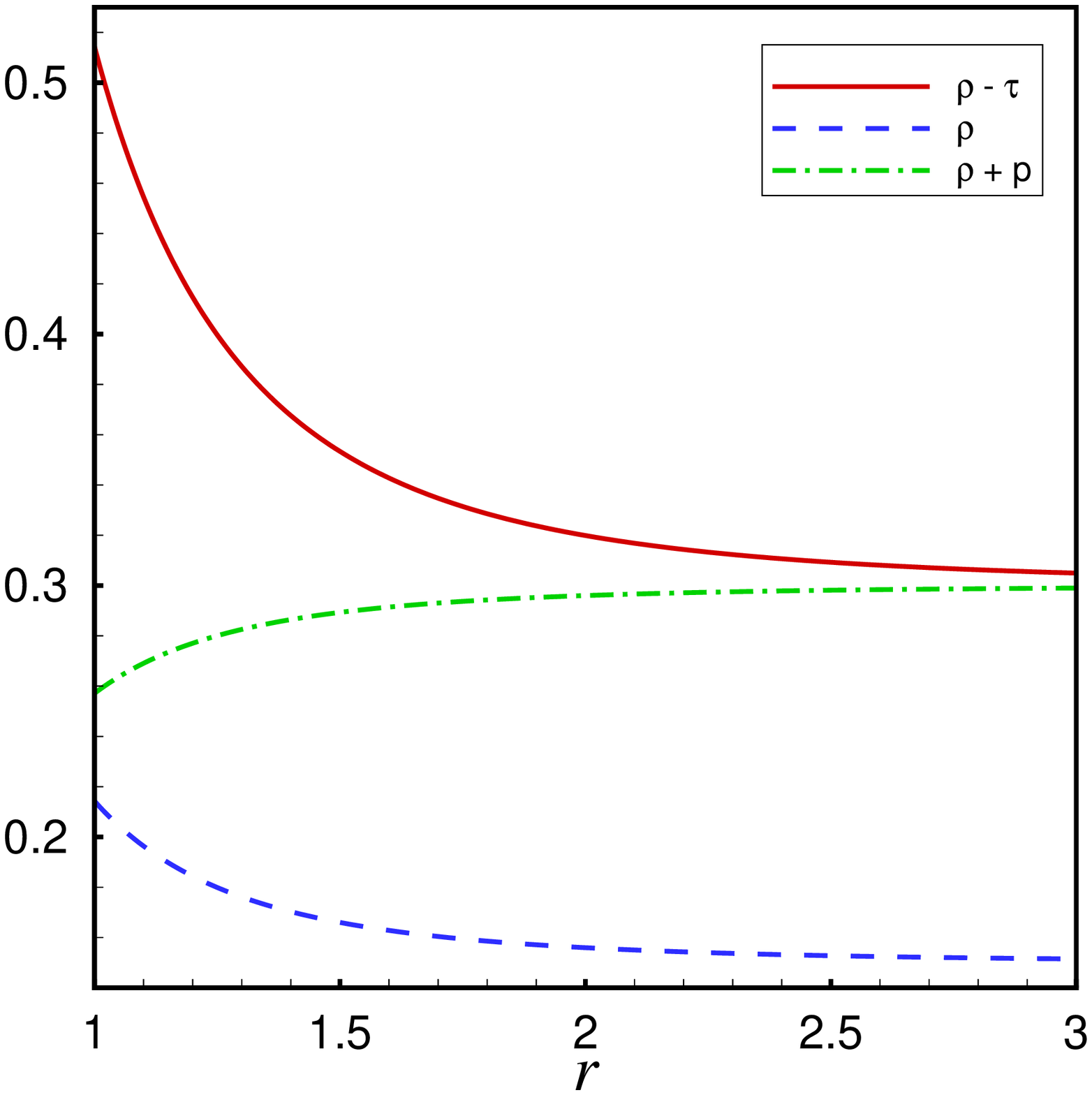}}
\caption{Left: $d(r)$ and $F(r)$ versus $r$; Right: $\protect\rho $, $%
\protect\rho -\protect\tau $ and $\protect\rho +p$ versus $r$ for $n=8$, $%
a_{0}=1$, $r_{0}=0.001$, $\protect\omega =1$, $k=-1$, $\protect\gamma =0.5$
and $\hat{k}=1$. Note that the scale of the vertical axes of both figures is
divided by $10^{7}$.}
\label{f2}
\end{figure*}

The conditions (\ref{Cond1}) and (\ref{Cond2}) for the general wormhole
solution (\ref{shape}) take the form 
\begin{eqnarray}
d(r) &=&\left( \hat{k}-a_{0}^{2}\beta \right) a_{0}^{1-\eta }r^{\eta }+\beta
r^{3}-\hat{k}r\leq 0,  \label{travers1} \\
F(r) &=&(\hat{k}-a_{0}^{2}\beta )(\eta -1)a_{0}^{1-\eta }r^{\eta }+2\beta
r^{3}<0,  \label{travers2}
\end{eqnarray}%
respectively. For $\hat{k}=0$, the above two conditions reduce to%
\begin{equation}
\beta \left[ 1-\left( \frac{r}{a_{0}}\right) ^{\eta -3}\right] \leq 0,
\end{equation}%
\begin{equation}
\beta \left[ 1-\frac{\eta -1}{2}\left( \frac{r}{a_{0}}\right) ^{\eta -3}%
\right] <0,
\end{equation}%
which are satisfied if 
\begin{equation}
\beta >0\text{ and \ }\eta \geq 3,\qquad \text{or}\qquad \beta <0\quad \text{%
and}\quad \eta \leq 3.  \label{khat0}
\end{equation}%
The intersection of the two conditions (\ref{wec}) and (\ref{khat0}) is
simply condition (\ref{wec}). Note that condition (\ref{wec}) in terms of $%
\omega $\ and $\gamma $ is given by the inequalities (\ref{wec1}) and (\ref%
{wec2}).

Since $a_{0}$ is the root of $d(r)$, one may discuss the negativity of $d(r)$
in the case of $\hat{k}=\pm 1$ by expanding $d(r)$ around $r=a_{0}$. Thus,
one obtains 
\begin{equation}
d(r)\left\vert _{r=a_{0}+\epsilon }\right. \approx \epsilon \left[ \beta
a_{0}^{2}(3-\eta )+\hat{k}(\eta -1)\right] ,  \label{trav1}
\end{equation}%
where $\epsilon $ is an infinitesimal positive parameter. Using Eq. (\ref%
{trav1}), condition (\ref{Cond1}) reduces to%
\begin{equation}
\beta a_{0}^{2}(3-\eta )+\hat{k}(\eta -1)\leq 0  \notag
\end{equation}%
or equivalently 
\begin{equation}
2a_{0}^{2}\beta +(\eta -1)(\hat{k}-a_{0}^{2}\beta )\leq 0,  \label{tra1}
\end{equation}%
which may be satisfied for specific choice of $a_{0}$ and $r_{0}$. More
specifically, the conditions under which (\ref{tra1}) is satisfied for
arbitrary values of $a_{0}$ and $r_{0}$ are given by: 
\begin{eqnarray}
\text{for \ \ \ \ }\hat{k} &=&1\Rightarrow \beta <0\text{ \ and \ }\eta \leq
1,  \notag \\
\text{for \ \ \ \ }\hat{k} &=&-1\Rightarrow \left\{ 
\begin{array}{l}
\beta >0\text{ \ and \ }\eta \geq 3 \\ 
\beta <0\text{ \ and \ }1\leq \eta \leq 3%
\end{array}%
\right. .  \label{first cond}
\end{eqnarray}%
Using (\ref{tra1}), one deduces the following inequality: 
\begin{equation}
F(r)=(\hat{k}-a_{0}^{2}\beta )(\eta -1)a_{0}^{1-\eta }r^{\eta }+2\beta
r^{3}\leq 2\beta (r^{3}-a_{0}^{3-\eta }r^{\eta }).  \label{tra2}
\end{equation}%
The parameter $a_{0}$ is the root of rhs of inequality (\ref{tra2}) and the
condition (\ref{Cond2}) is satisfied provided the rhs is negative%
\begin{equation}
\text{rhs}\left\vert _{r=a_{0}+\epsilon }\right. \approx 2a_{0}^{2}\epsilon
\beta (3-\eta )\leq 0,
\end{equation}%
which is satisfied if%
\begin{equation}
\beta >0\quad \text{and}\quad \eta \geq 3,\qquad \text{or}\qquad \beta
<0\quad \text{and}\quad \eta \leq 3.  \label{sec cond}
\end{equation}%
Note that the intersection of the two sets (\ref{sec cond}) and (\ref{first
cond}) is simply (\ref{first cond}). Therefore, the conditions (\ref{Cond1})
and (\ref{Cond2}) are satisfied if and only if (\ref{first cond}) are
satisfied. Furthermore, the intersection of the two sets (\ref{wec}) and (%
\ref{first cond}) is condition (\ref{wec}). Thus, conditions (\ref{Cond1})
and (\ref{Cond2}), and the energy conditions are satisfied if and only if (%
\ref{wec}) is satisfied. (As mentioned before, (\ref{wec}) is expressed in
terms of $\omega $ and $\gamma $ by (\ref{wec1}) and (\ref{wec2})).

The above discussions are depicted in Figs. (\ref{f1}) and (\ref{f2}). The
left hand side plots of Figs. (\ref{f1}) and (\ref{f2}) show the negativity
of the functions $d(r)$ and $F(r)$ for two appropriate settings of
parameters. The right hand side plots show that for these sets of parameters
the WEC is also satisfied. Therefore, it is shown that by considering
appropriate parameters, it is possible to construct a wormhole with
physically reasonable properties and with normal matter.

\subsection{Spherically-symmetric wormholes in an Ads universe}

\label{sphwo}

In this subsection, we consider a wormhole solution in the spacetime $r\in
\lbrack a_{0},a)$. In the exterior region $r>a$, the metric should satisfy
the field equation $\mathcal{G}_{\hat{a}\hat{b}\ }=0$, where $\mathcal{G}_{%
\hat{a}\hat{b}\ }$ is given in Eq. (\ref{Gab2}). Thus, the metric of a
spherically symmetric spacetime for $r>a$ may be written as \cite{diffpo4}%
\begin{equation}
ds^{2}=-f(r)dt^{2}+\frac{dr^{2}}{f(r)}+r^{2}d\Sigma _{2(\hat{k})}^{2},
\label{ds}
\end{equation}%
where%
\begin{equation}
f(r)=\hat{k}-\frac{2m}{r}-\frac{\Lambda _{\mathrm{eff}}}{3}r^{2}.
\end{equation}%
The background shows an asymptotically de--Sitter (dS) or anti de-Sitter
(AdS) spacetime for $k=1$ or $k=-1$, respectively (note that $\Lambda _{%
\mathrm{eff}}=(2n-13)k/r_{0}^{2}$). Using the wormhole metric (\ref{metric}%
), the continuity of the metric at the boundary $r=a$\ leads to%
\begin{eqnarray}
f(a) &=&e^{2\Phi } ,  \label{junc2} \\
b(a) &=&a-ae^{2\Phi }.  \label{junc1}
\end{eqnarray}

In the case of $\hat{k}=1$, by setting $\Phi =0$ one has 
\begin{equation*}
\frac{2m}{a}+\frac{\Lambda _{\mathrm{eff}}}{3}a^{2}=0 \qquad \Longrightarrow
\qquad a=\left(- \frac{6m}{\Lambda _{\mathrm{eff}}}\right) ^{\frac{1}{3}},
\end{equation*}%
and $b(a)=0$, where $m$\ may be set equal to the mass inside the region $%
r\in \lbrack a_{0},a)$, which can be calculated by using Eq. (\ref{rhor}) as%
\begin{eqnarray}
m &=&\int_{a_{0}}^{a=\left( \frac{-6m}{\Lambda _{\mathrm{eff}}}\right) ^{%
\frac{1}{3}}}4\pi \rho (r)r^{2}dr  \notag \\
&=&-\frac{32\pi }{\Xi }\left[ \frac{a_{0}^{3}\Lambda _{\mathrm{eff}}}{3}%
+2m-a_{0}\right] .  \label{mass}
\end{eqnarray}%
Note that in above calculations we do not use any specific solution for $%
b(r) $, namely, in order to determine Eq. (\ref{mass}), only the conditions $%
b(a)=0$ and $b(a_{0})=a_{0}$ were used. Using Eq. (\ref{mass}), $m$ can be
calculated as%
\begin{eqnarray}
m &=&\frac{32\pi a_{0}}{\Xi +64\pi }\left[ 1-\frac{a_{0}^{2}\Lambda _{%
\mathrm{eff}}}{3}\right]  \notag \\
&=&\frac{32\pi a_{0}}{\Xi +64\pi }\left[ 1-\frac{a_{0}^{2}(2n-13)k}{%
3r_{0}^{2}}\right]
\end{eqnarray}%
Since $a$($=\left( -6m/\Lambda _{\mathrm{eff}}\right) ^{1/3}$) and $m$
should be positive, one can find that (note $n\geq 7$)%
\begin{equation*}
\Lambda _{\mathrm{eff}}<0 \quad \text{ or equivalently } \quad k=-1.
\end{equation*}%
Therefore $m$ is given by 
\begin{equation*}
m=\frac{32\pi a_{0}}{\Xi +64\pi }\left[ 1+\frac{a_{0}^{2}(2n-13)}{3r_{0}^{2}}%
\right] >0 .
\end{equation*}%
It is remarkable to note that the possibility of an accelerating expansion
in theories with negative cosmological constant has been shown recently \cite%
{haw}.

\section{Summary and Discussion}

In this paper, we first decomposed the field equations of third order
Lovelock gravity in the presence of cosmological constant and matter field
in an $n$-dimensional spacetime, $\mathcal{M}^{4}\times \mathcal{K}^{n-4}$,
into two gravitational field equations. We fixed the Lovelock coefficients
and the cosmological constant in terms of the size of the extra dimensions $%
r_{0}$ and the dimension of the spacetime by applying the idea that the $%
\mathcal{K}^{n-4}$ space is empty. This idea also implies that $n\geq 7$.

Furthermore, the energy density, the radial and tangential pressures
threading the wormhole solution were obtained using fixed Lovelock
coefficients and $\Lambda $. As an interesting example, the equation of
state $\rho =(\gamma p-\tau )/[\omega (1+\gamma )]$ \cite{EOS} was used to
find the exact solution for the shape function in the case of a zero tidal
force, i.e., of a constant redshift function. Next, the energy conditions
were substantially investigated and the conditions under which the normal
matter could sustain a wormhole structure were analysed. Then, physical
properties and characteristics of the wormhole solutions were investigated
by imposing specific conditions on the shape function. Indeed, it was shown
that it is possible to have asymptotically flat and non-flat traversable
wormhole structure satisfying the WEC (see Figs. \ref{f1} and \ref{f2}). We
considered a limited wormhole in a $4$-dimensional asymptotically non-flat
spacetime, and found the relationship between the wormhole parameters and
the cosmological constant corresponding to the $4$-dimensional spacetime. It
was shown that if a spherically symmetric wormhole is constructed with
normal matter, this implies a negative cosmological constant. In this
context, it is worth mentioning that the possibility of accelerating
expansion in the presence of a negative cosmological constant has been shown
recently \cite{haw}.

In the analysis outlined in this paper, we have followed the second order
formalism, by assuming torsion to be zero. However, in the first order
approach, where the independent dynamical variables are the vielbein and the
spin connection, which obey first order differential field equations, the
equations of motion do not imply the vanishing of torsion, which is a
propagating degree of freedom \cite{Troncoso:1999pk}. In addition to this,
an advantage of the first order formalism is that it can be written out
entirely in terms of differential forms and their exterior derivatives,
without the introduction of the inverse vielbein. Several exact solutions
with non-trivial torsion have been extensively investigated in the
literature \cite{1storderexact,Canfora:2008ka,Anabalon:2011bw}. More
specifically, in Lovelock gravity using the first order formalism, the
equations of motion do not imply the vanishing of torsion in vacuum as in
general relativity, so that torsion may also have propagating degrees of
freedom, as mentioned above. Thus, an interesting future project would be to
extend the analysis carried out in this work, in the first order formalism,
by considering the approach outlined in \cite{Anabalon:2011bw}.

The latter approach is particularly interesting in light of the vacuum
static compactified wormholes found in $8$-dimensional Lovelock theory \cite%
{Canfora:2008ka}. Indeed, the exact static vacuum $8$-dimensional solutions
were constructed with the structure of $M_{5}\times \Sigma _{3}$, where $%
M_{5}\equiv M_{2}\times F(r)N_{3}$ is a $5$-dimensional manifold, where $%
M_{2}$ plays the role of an $r-t$ plane, $N_{3}$ is a constant curvature
manifold, $F(r)$ is a warp factor, and $\Sigma _{3}$ is a compact constant
curvature manifold that plays the role of a compactified $3$-dimensioanl
space. The role of torsion is of particular interest as it was shown that
these wormholes solutions with torsion act as a geometrical filter, which
may distinguish between scalars and spinors and also between the helicities
of particles. More specifically, it was also shown that a very large torsion
may increase the traversability properties for scalars and a
\textquotedblleft polarizator\textquotedblright\ on spinning particles. In
comparing these latter results with those outlined in the present paper, we
have been essentially interested in the analysis of the energy conditions,
and our results differ radcially form those found in \cite{Canfora:2008ka}.
However, it is of interest to extend our analysis, in thrid-order Lovelock
theory, to include the effects of torsion, and not only investigate the
traversability conditions but to analyse the energy conditions. Work along
these lines is presently under consideration.

An interesting and important issue is related to the stability of these
solutions. One may ask which family of compactified wormholes is more
stable? Indeed, the stability issue in Lovelock theory is very complicated
and certainly lies outside the scope of this paper. However, as a first
approach one may analyze just the Euclidean actions of the different
configurations. At lowest order, the Euclidean action provides the free
energy and so one could compare the novel solutions presented in this work
with some of the already known solutions by comparing the corresponding
Euclidean actions. In this way, one could find the range of parameters in
which the novel solutions are favored with respect to previous ones.
However, this stabiltiy analysis lies completely outside the scope of this
paper and will be addressed in future work, in addition to a full-fledged
stabililty analysis. Work along these lines are currently underway.

\acknowledgments
We thank an anonymous referee for an extremely constructive report, and for
helpful suggestions and comments that have significantly improved the paper.
This work has been supported by Research Institute for Astrophysics and
Astronomy of Maragha. FSNL acknowledges financial support of the Funda\c{c}%
\~{a}o para a Ci\^{e}ncia e Tecnologia through an Investigador FCT Research
contract, with reference IF/00859/2012, and the grants
PEst-OE/FIS/UI2751/2014 and UID/FIS/04434/2013.

\end{document}